\documentclass[aps,pra,twocolumn,showpacs,superscriptaddress,preprintnumbers,amsmath,amssymb,nofootinbib]{revtex4}

\usepackage{graphicx}
\usepackage{mathtools}
\usepackage{gensymb}
\begin{document}

\title{Homogenization of Maxwell's equations in layered system beyond static approximation}

\author{R.S. Puzko}
\email[]{roman998@mail.ru}
\affiliation{All-Russia Research Institute of Automatics, 22, ul. Sushchevskaya, Moscow 127055, Russia}
\affiliation{Moscow Institute of Physics and Technology, 9 Institutskiy per., Dolgoprudny, Moscow Region, 141700, Russia}

\author{A.M. Merzlikin}
\email[]{merzlikin_a@mail.ru}
\affiliation{All-Russia Research Institute of Automatics, 22, ul. Sushchevskaya, Moscow 127055, Russia}
\affiliation{Moscow Institute of Physics and Technology, 9 Institutskiy per., Dolgoprudny, Moscow Region, 141700, Russia}
\affiliation{Institute for Theoretical and Applied Electromagnetics Russian Academy of Sciences, 13, ul. Izhorskaya, Moscow 125412, Russia}

\date{\today}

\begin{abstract}
The propagation of electromagnetic waves through disordered layered system is considered in the paradigm of Maxwell's equations homogenization. In spite of the impossibility to describe the system in terms of effective dielectric permittivity and/or magnetic permeability the unified way to describe the propagation and Anderson localization of electromagnetic waves is proposed in terms of the introduced effective wave vector (effective refractive index). It is demonstrated that both real and imaginary parts of the effective wave vector (contrary to effective dielectric permittivity and/or magnetic permeability) are the self-averaging quantities at any frequency. The introduced effective wave vector is analytical function of frequency; corresponding Kramers-Kronig relation generalizes the Jones-Herbert-Thouless formula.
\end{abstract}

\maketitle

%%%%%%%%%%%%%%%%%%%%%%%%%%  body  %%%%%%%%%%%%%%%%%%%%%%%%%%
\section{Introduction}
\begin{figure}
\centering\includegraphics[width=1\linewidth]{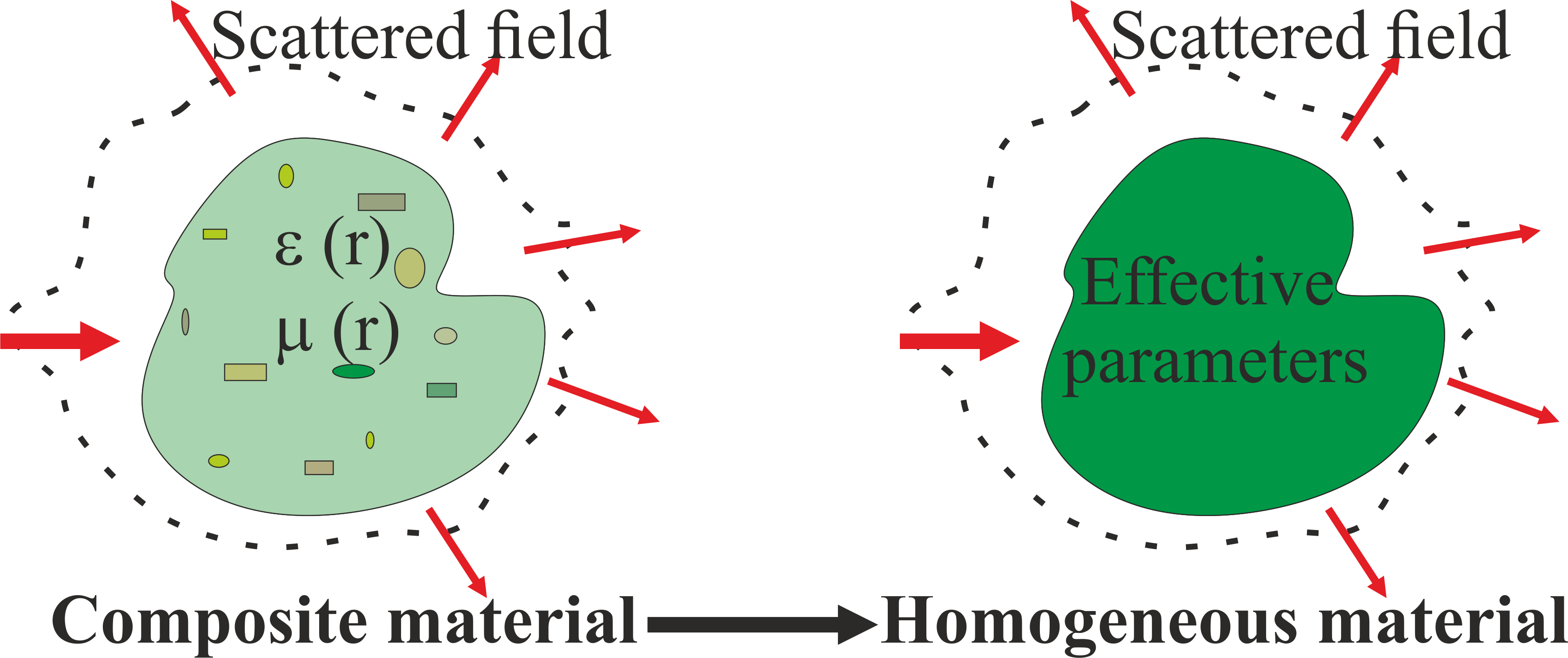}
\caption{Homogenization problem.\label{Fig0}}
\end{figure}

The electrodynamic properties of artificial composite materials depend on their structure and the components compounds, and can differ significantly from the properties of homogeneous materials \cite{metamaterials}. Nevertheless, the description of metamaterials as homogeneous materials by means of effective parameters (dielectric permittivity and magnetic permeability, chirality coefficients, etc.), still remains one of the relevant problems of electrodynamics. The problem of replacing an inhomogeneous composite system by a homogeneous material having effective parameters is called the homogenization problem (see Fig.~\ref{Fig0}). The parameters of the homogeneous system provide that the scattered field is the same in both cases and called "effective parameters". I.e. the effective parameters allow describing light scattering by composite system without considering system inhomogeneous structure.

The general formulation of homogenization problem inside the area filled with inhomogeneous composite system implies~\cite{vinogradov2012comment}:

1. Derivation of equations describing the macroscopic fields inside the area (these equations and macroscopic fields may completely differ from the Maxwell's equations and electric and magnetic fields). The derived equations imply parameters which are effective material parameters. Effective material parameters depend only on microscopic structure of composite material (density of inclusions, inclusions shape distribution function, correlation functions etc.) and are independent on the area shape\footnote{As soon as material parameters should not depend on composite system shape (and depend on composite microstructure), the composite system has to be large enough to apply the statistical description.}. Derivation of boundary conditions, i.e. some relations between the macroscopic fields and the external electric and magnetic fields.

2. The area with inhomogeneous system is replaced by the area filled by homogeneous material having effective parameters. The derived equations, effective material parameters and boundary conditions should provide that the difference (in terms of fields) between the two cases is zero or negligibly small\footnote{As a rule, in the limit $\xi<<\lambda$ the difference should be of order less than $(\xi/\lambda)^2$, where $\xi$ is inhomogeneity size scale} outside the area.

As a rule, it is assumed that the linear size $L$ of the composite system is so large, and the characteristic size of the inhomogeneities is so small that the composite can be considered as homogeneous in the macroscopic sense (obviously, the description of the two-layer system by effective parameters looks curiously). In addition, if we have some procedure providing us with the effective parameters for a finite sample of size $L$, then, as $L$ increases, these parameters should tend to some constant values. This property (the tendency to a constant with an increase in the system size) of the effective parameter of the random system is called "self-averaging".

It should be pointed out that homogenization problems for cases of electrostatics and electrodynamics are significantly different. In the static case, Maxwell's equations are divided into two groups of equations - magnetic and electric. Therefore, the description of inhomogeneous media by means of effective dielectric permittivity $\varepsilon_{eff}$ and magnetic permeability $\mu_{eff}$ is quite natural. The microscopic description of each of the fields (electric and magnetic) reduces to a dimensionless Laplace equation, and scaling approach can be applied \cite{sanchez1980nonhomogeneous}. At the moment, the greatest progress was made in homogenizing the Maxwell's equations for the static case. Namely, the scaling theory of G-convergence \cite{sanchez1980nonhomogeneous} and spectral theory \cite{bergman1978dielectric, bergman1992physical,mcphedran1981bounds,milton1981bounds} were created.

In contrast to statics, the electric and magnetic fields are related to each other in case of dynamics. In particular, the amplitudes of electric and magnetic fields of plane wave amplitudes are connected by impedance (admittance). In addition, the homogenization problem in dynamics becomes multiscale - the scales appear, which are associated with the equations, namely, the set of local wavelengths $\lambda_0/\sqrt{\varepsilon\mu}$, where $\lambda_0=2\pi/k_0$ is the wavelength in vacuum, $k_0=\omega/c$ is the wave number. The presence of many scales leads to certain difficulties in the transition from the finite system to the infinite one \cite{lagarkov_vinogradov2012advances}, and the results of the G-convergence theory cannot be applied in the general case. Thus, two questions arise when the theory of homogenization problem for the dynamics is being constructed: which parameters should be chosen as effective, and how the homogenization procedure should be carried out \cite{vinogradov2012comment,chipouline2012basics,tsukerman2014non,krokhin2016high}?
 
Currently, the theory of homogenization problem is built only for the static case - it is mentioned above spectral theory (a special case of which is G-convergence). Outside of static case the theory of homogenization problem is still not established (even in the quasi-static approximation). The existing approaches to homogenization problem in dynamics predict different results.

One of the approaches to homogenization problem was proposed for periodic systems in the works of Wood, Ashcroft and Datta \cite{lamb1980long,datta1993effective} and was recently developed in \cite{krokhin2002long,pendry1994photonic}. The basic idea is to determine the effective dielectric permittivity of an inhomogeneous periodic dielectric medium by means of Bloch wave vector $k_{Bl}=k_0\sqrt{\varepsilon_{eff}}$. That is, it is implied that in the absence of magnetic components, we can consider that $\mu_{eff}=1$. In essence, this approach is equivalent to the fact that we transfer the independence of magnetic and electric fields from statics to dynamics (that is, the effective permittivity and permeability are introduced by homogenization procedure independently). Particularly, the impedance is implied to be equal to the inverse refractive index. This is a common case in the optics of natural media. However, this approach has a disadvantage of poor description of metamaterials possessing the artificial magnetism.

An alternative approach was proposed by S.M. Rytov and M.L. Levin \cite{rytov1956electromagnetic,levin48,brekhovskikh1948waves,rouhani1986effective,akcakaya1988effective,kikarin1989effective,semchenko1990gyrotropic} for layered systems and was generalized to the two-dimensional and three-dimensional case \cite{smith2006homogenization,acher2007evaluation,silveirinha2010casimir,cerdan2009anisotropy,yakovlev2016local,chebykin2015spatial,yagupov2015diamagnetism}. The basic idea of all approaches of that kind is the averaging of the fields over the unit cell of the periodic medium. The profound variation of the field averaging approach considers finite systems~\cite{andryieuski2009}, i.e. the introduced effective admittance directly accounts the reflection from the finite sample. In the latter case the averaging procedure is carried out over the unit cell which is far enough from the system boundaries to avoid the boundary effects contribution.

\begin{figure}
\centering\includegraphics[width=0.9\linewidth]{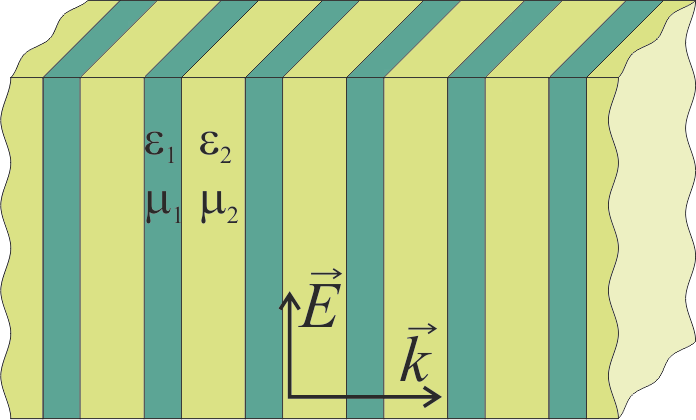}
\caption{Periodic layered system. The unit cell consists of layers ($d_1$, $\varepsilon_1$, $\mu_1$) and ($d_2$, $\varepsilon_2$, $\mu_2$).\label{Rytov_system}}
\end{figure}

Let us consider this alternative approach in more detail by the example of a periodic layered system where the plane wave is propagating perpendicularly to the layers. The cell of the considered system consists of two layers: the thicknesses of the layers are $d_1$ and $d_2$, their dielectric permittivities are $\varepsilon_1$ and $\varepsilon_2$, the magnetic permeabilities are $\mu_1$ and $\mu_2$ (see Fig.~\ref{Rytov_system}).

Within this approach it is proposed to take the Bloch wave vector $k_{Bl}$ as the effective wave vector $k_{eff}^{Ryt}$ (and the corresponding effective refractive index is $n_{eff}^{Ryt}=\sqrt{\varepsilon_{eff}^{Ryt}\mu_{eff}^{Ryt}}=k_{eff}^{Ryt}/k_0$). It is important that the knowledge of the effective refractive index is not enough for the complete description of the environment (for example, finding both $\varepsilon_{eff}$ and $\mu_{eff}$): it is also necessary to determine the effective admittance $Y_{eff}^{Ryt}=\sqrt{\varepsilon_{eff}^{Ryt}/\mu_{eff}^{Ryt}}$, which was defined as $Y_{eff}^{Ryt}=\langle H \rangle /\langle E \rangle$, where the averaging of the fields is taken over the unit cell. There are alternative definitions of $Y_{eff}$, which imply the two posible way of field averaging: over the unit cell or over the surface~\cite{andryieuski2012}. However, the corrections to static formulas lead to a somewhat dubious result: either $\varepsilon_{eff}$ or $\mu_{eff}$ has a negative imaginary part.

In order to handle this unexpected consequence, it is necessary to consider the finite system and compare the mentioned above parameters (defined in alternative Rytov's approach) with the effective parameters directly retrieved for that system from the scattering data \cite{vinogradov2004band,puzko2017analytical}. In the case of directly retrieved parameters, the values of $\varepsilon_{eff}$ and $\mu_{eff}$ are determined by means of the transmission $t$ and the reflection $r$ coefficients of the incident wave \cite{smith2002determination,lagarkov1998dielectric,mota2016constitutive}
\begin{eqnarray}
	\label{Ykeff}
	Y_{eff}=\sqrt{\frac{(1-r)^2-t^2}{(1+r)^2-t^2}},\label{Y_eff_Soukalis}\\
	e^{ik_{eff}L}=\frac{t(1+Y_{eff})}{Y_{eff}+rY_{eff}+1-r}\label{k_eff_Soukalis},
\end{eqnarray}
and, correspondingly, $\mu_{eff}=k_{eff}/(Y_{eff}k_0)$ and $\varepsilon_{eff}=k_{eff}Y_{eff}/k_0$.

\begin{figure}
	\begin{minipage}{0.49\linewidth}
		\centering\includegraphics[width=1.0\linewidth]{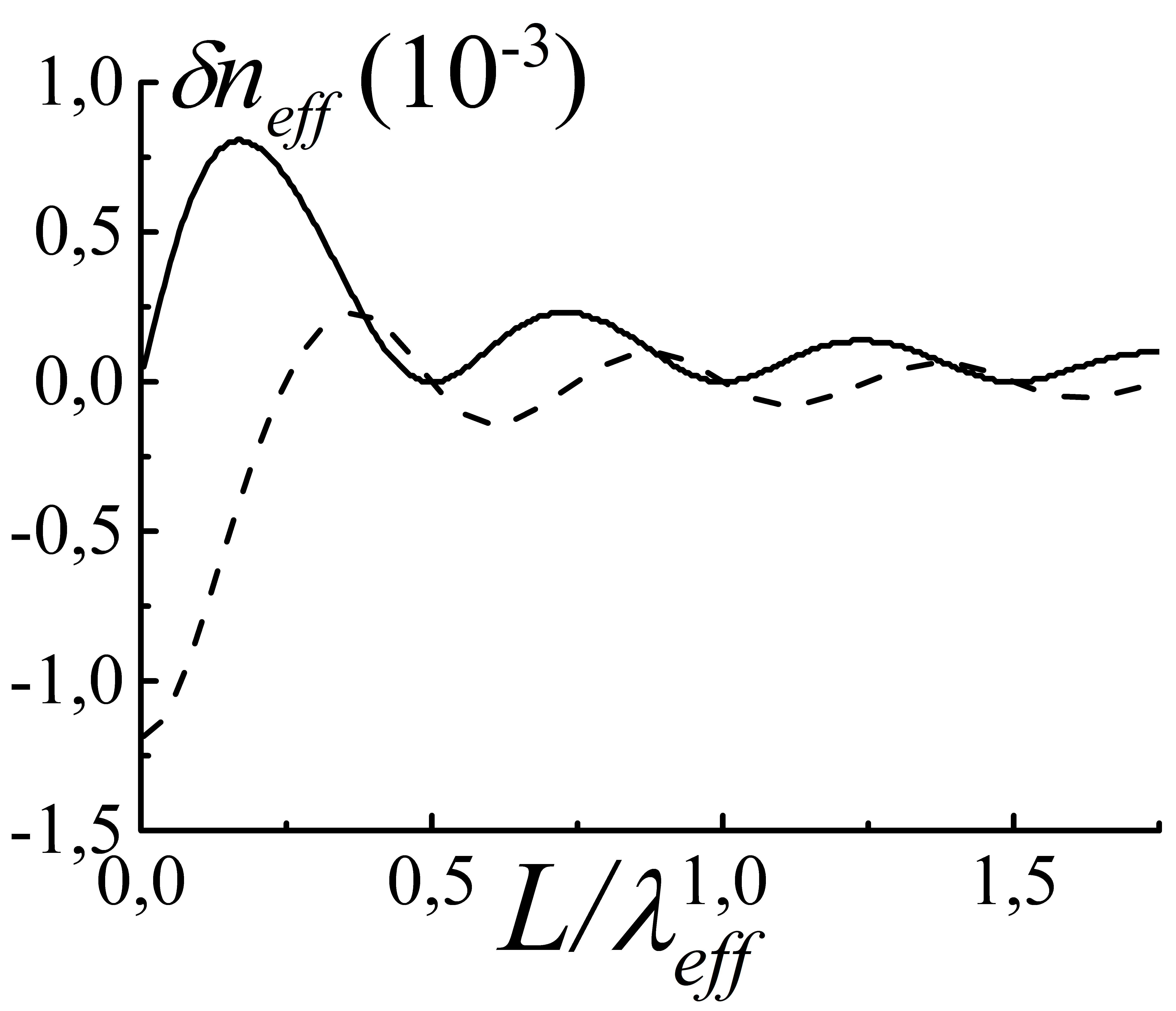}
		(a)
	\end{minipage}
	\begin{minipage}{0.49\linewidth}
		\centering\includegraphics[width=1.0\linewidth]{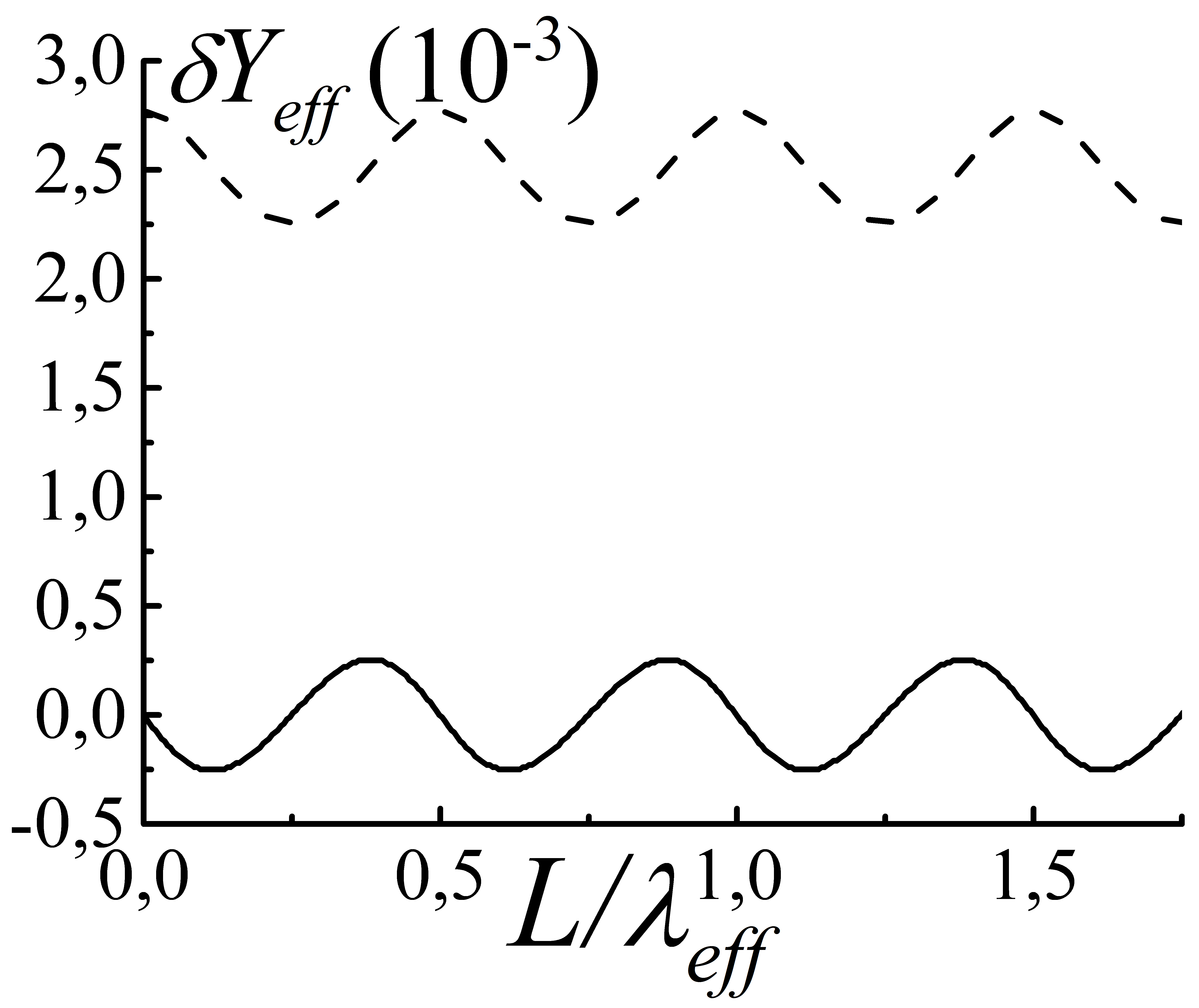} 
		(b)
	\end{minipage}
	\caption{The dependencies of (a) $\delta n=n_{eff}-n_{eff}^{Ryt}$ and (b) $\delta Y=Y_{eff}-Y_{eff}^{Ryt}$ on the thickness of the periodic system in case of even number of layers: real and imaginary parts of the quantities displayed by solid and dotted curves, respectively. Parameters of the layers of the unit cell: $\varepsilon_1=2$, $\varepsilon_2=3$, $k_0d_1=k_0d_2=0.01$.\label{n_Y_eff}}
\end{figure}

\begin{figure}
	\begin{minipage}{0.49\linewidth}
		\centering\includegraphics[width=1\linewidth]{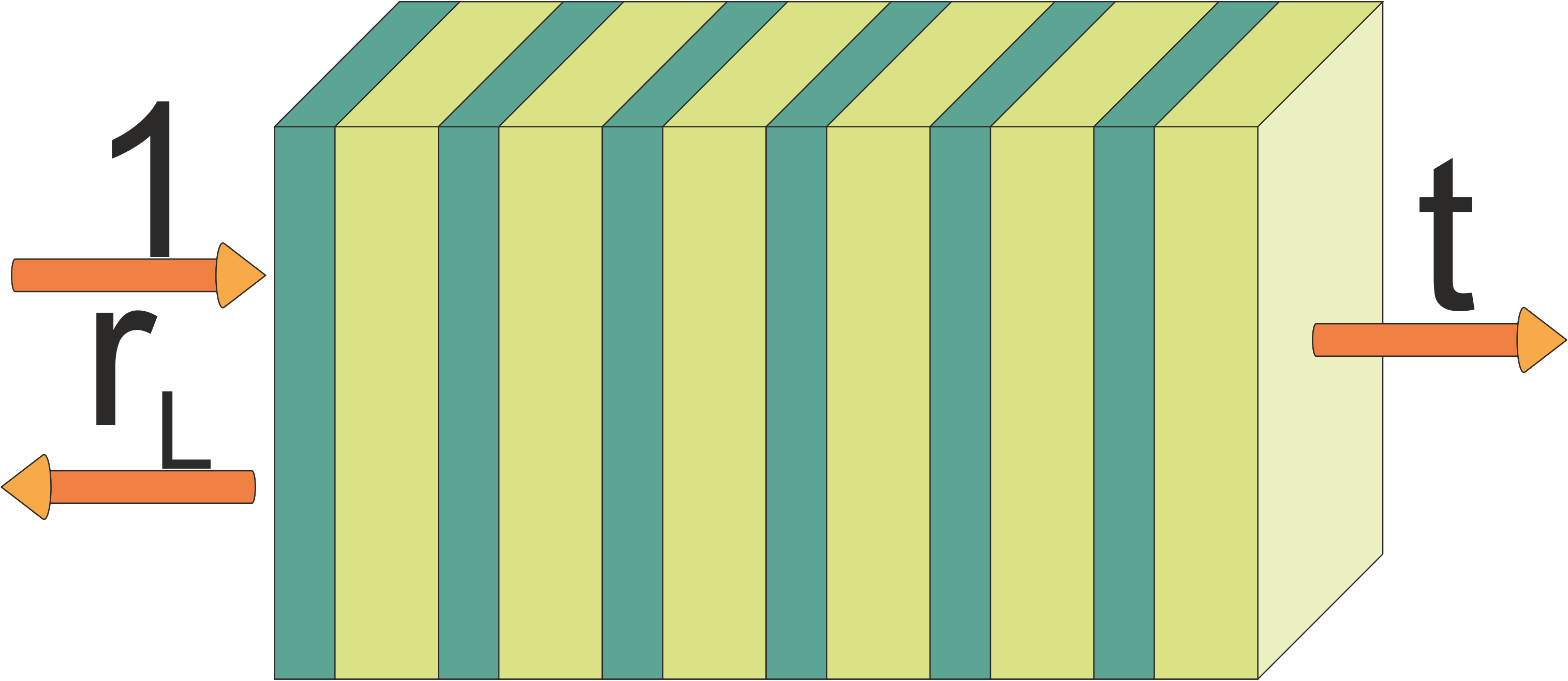}
		(a)
	\end{minipage}
	\begin{minipage}{0.49\linewidth}
		\centering\includegraphics[width=1\linewidth]{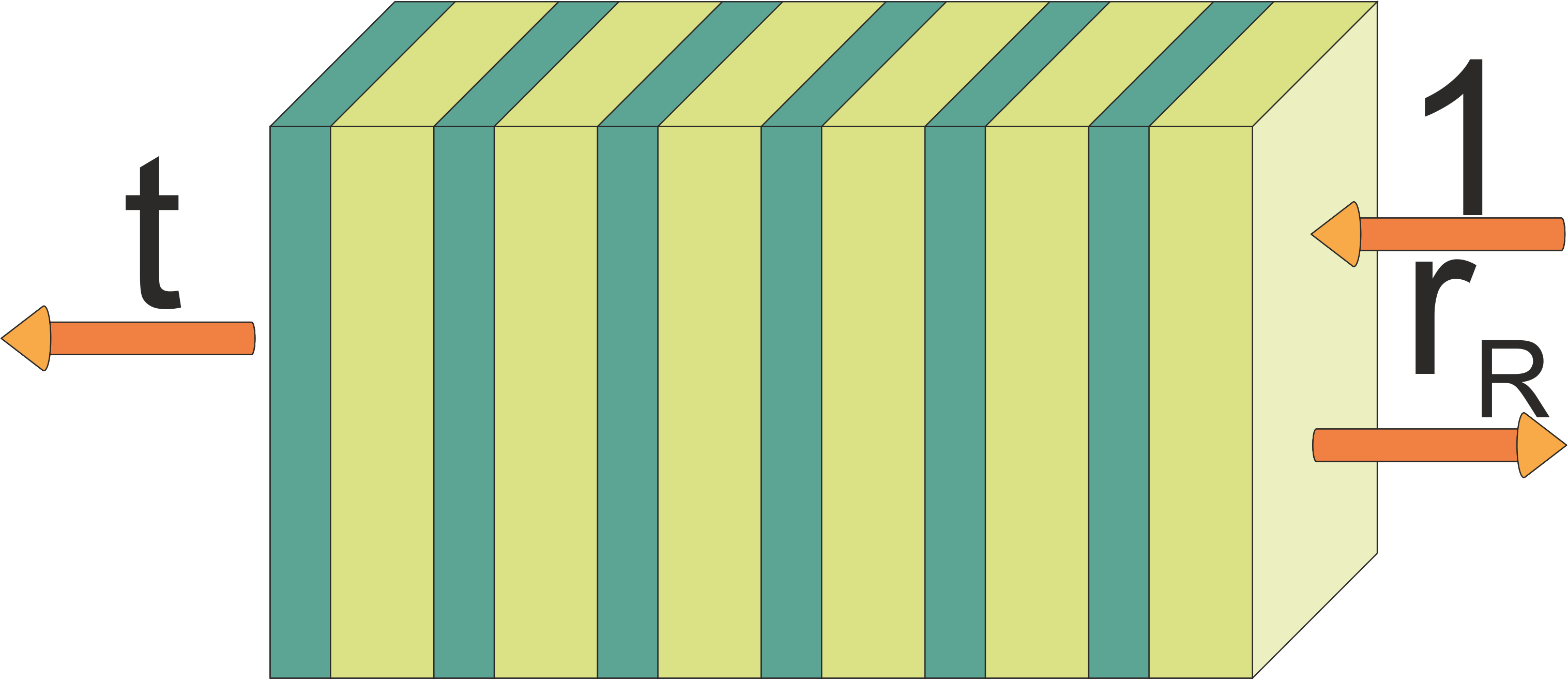}
		(b)
	\end{minipage}
	\caption{Light incidence from the left (a) and from the right (b) to the system.\label{left_right}}
\end{figure}

Let us consider the system composed of arbitrary integer number of cells, that is, the number of layers is even (the case of odd number of layers is similar \cite{vinogradov2002problem,vinogradov2001electrodynamic}). Calculations of (\ref{Ykeff}) in that case show that as the size of the system increases\footnote{The methods of $Y_{eff}$ introduction by means of an additional layer or surface currents are proposed in a number of works \cite{vinogradov2011additional,simovski2007application}. These methods allow escaping the dependency of $Y_{eff}$ on the number of layers.}, $n_{eff}=k_{eff}/k_0$  tends to $n_{eff}^{Ryt}=k_{eff}^{Ryt}/k_0$ , and $Y_{eff}$ oscillates with a period $\lambda_{eff}/2$ (Fig.~\ref{n_Y_eff}). The quantities $\varepsilon_{eff}$ and $\mu_{eff}$ which are the functions of both $n_{eff}$ and $Y_{eff}$ show the combined behavior, namely, being asymptotically periodic functions, they do not tend to any limit at $L\to\infty$.
Thus, $\varepsilon_{eff}$ and $\mu_{eff}$ are not self-averaging quantities for layered systems outside the static approximation, so their introduction might be incorrect. This fact can be easily explained. Indeed, let us come back to a finite periodic system (the unit cell of which consists of two layers) composed of an even number of layers (see Fig.~\ref{left_right}). The first and the last layers are made of different materials (the sample is not symmetrical with respect to the direction of light propagation). The calculations show that the amplitude reflection coefficients from the different sides of the system $r_L$ and $r_R$ are different (see Fig.~\ref{left_right}). That contradicts the case of homogeneous layer, for which the equality $r_L=r_R$ takes place for any dielectric permittivity and magnetic permeability. Direct calculations show that the difference between the coefficients $r_L$ and $r_R$ is of the first order of $k_0d$. The phases are different even in the absence of losses. The problem can be partially solved by considering homogenization procedure of higher order in $k_0d$. The second-order homogenization procedure describes chirality, magnetic dipoles and electric quadrupoles  and therefore allows to avoid the contradictions in the longwave approximation~\cite{popov2016}. However, such effective medium theory is still experience the breakdown as the parameter $k_0d$ increases. Moreover, the validity of the theory depends on the system length~\cite{popov2016}. That fact leads to the conclusion that the homogenization procedure in terms of effective $\varepsilon_{eff}$ and $\mu_{eff}$ is possible only in statics ($\lambda\to\infty$) \cite{vinogradov2001electrodynamic,vinogradov2002electromagnetic,zouhdi2012advances}.

It is necessary to mention another approach to the homogenization problem \cite{tsukerman2014non}, whose basic idea is to approximate the precise solution of Maxwell's equations for a sample. The approximated electromagnetic fields are used then to introduce the effective material parameters. Despite the fact that the introduced effective parameters lead to some homogenization errors, the errors can be minimized by applying certain optimization procedures \cite{tsukerman2014non}.

It should be pointed out that the problem of effective parameters introduction can be formulated not only in terms of $\varepsilon_{eff}$ and $\mu_{eff}$. Even though they do not experience self-averaging (and therefore they describe a certain finite sample rather than a material), it was shown in \cite{vinogradov2002problem,puzko2017analytical} that the effective refractive index self-averages (see Fig.~\ref{n_Y_eff}). Thus, the homogenization problem beyond the static approximation reduces to two independent problems: determining the effective refractive index and considering the boundary conditions. To solve the problem of boundary conditions, the regularization of the effective impedance was earlier proposed by means of additional surface currents \cite{vinogradov2011additional} or an additional effective layer at the boundary \cite{simovski2007application}. In this paper, we distract from the boundary effects and focus our attention on the properties of $n_{eff}$.

In the simplest case of a periodic system, calculations show \cite{puzko2017analytical} that using only the transmission coefficient, it is possible to introduce an effective wave vector that converges to the Bloch vector as the number of layers of the system increases. Moreover, convergence is observed not only in the long-wavelength approximation \cite{vinogradov2002problem}, but also for any ratio of the wavelength to the characteristic thickness of the layers \cite{puzko2017analytical}. The derivative of the introduced effective wave vector has the property of analyticity and, consequently, satisfies Kramers-Kronig like relations. The Kramers-Kronig like relations are well-known for different optical parameters, in particular, complex refractive index~\cite{bookKramKron}. The relations are in close connection with the causality principle.

In this paper the properties of $k_{eff}$ mentioned above are considered in detail in connection with the disordered systems. In particular the relation between the localization length and the $k_{eff}$ is established.

\section{Self-averaging of the effective wave vector}
We study the wave propagation through a 1D disordered system composed of dielectric layers with different dielectric permittivities. For simplicity, we assume that all the layers are of the same thickness.

The standard approach to the localization and propagation of light considers a disordered system immersed in a vacuum. In this case, the transmission and reflection coefficients and other scattering data are affected both by the Anderson localization \cite{anderson1958absence} and by the effect of reflection from boundaries between the vacuum and the disordered system. In order to separate the localization effect from the scattering at the boundaries, we immerse the system under consideration into a medium with an averaged permittivity (the averaging is taken over the permittivity probability distribution).
The transmission coefficient of a one-dimensional disordered system can be represented in the following form:
\begin{equation}
t=|t|\mathrm{exp}\left(i\phi\right)=\mathrm{exp}\left(i\mathrm{Im}({\ln}t)+\ln{⁡|t|}\right).
\label{t_represent}
\end{equation}

\begin{figure}
	\centering\includegraphics[width=1\linewidth]{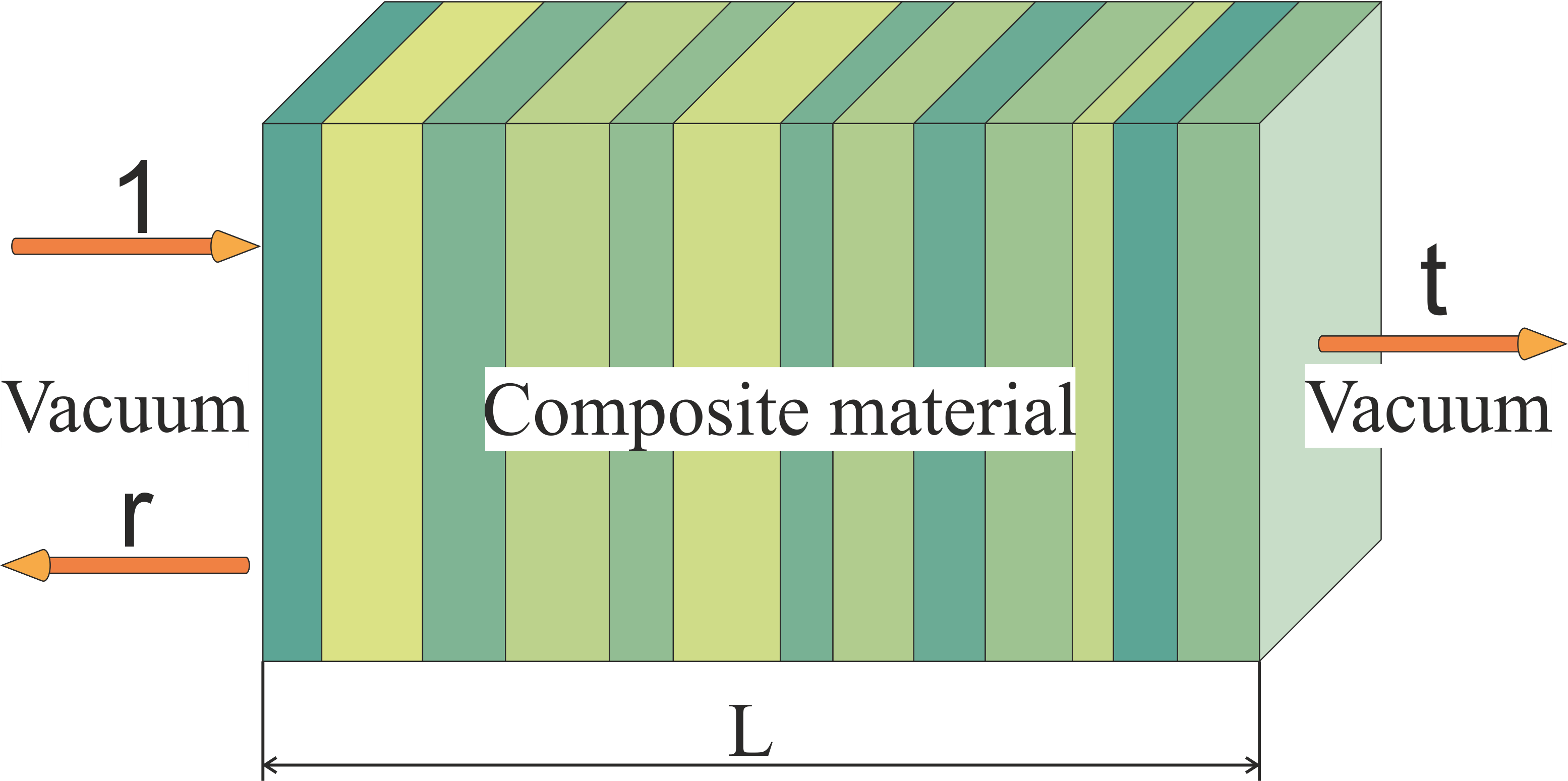}
	\caption{Propagation of a plane wave through a composite system.\label{random_system}}
\end{figure}

The propagation of the electromagnetic wave through the inhomogeneous layer can be viewed as a three-stage process: transmission of the vacuum-layer boundary, propagation through the volume of the layer, and transmission of the layer-vacuum boundary (see Fig.~\ref{random_system}).
It should be noted that the contribution of the boundaries of the system to the transmission coefficient does not increase (or increases more slowly than exponentially) as the system thickness increases. At the same time, propagation through the layer is characterized by an effective wave vector, whose contribution to the phase or exponential decay (associated with absorption or scattering of light) of the eigensolution increases linearly alongside with the thickness $L$ of the system. Thus, the reasonable approximation used in our discussion neglects the surface effects for large $L$ (that is, the phase is much larger than $\pi$, i.e. $L\gg\lambda$). Therefore, the imaginary part of the logarithm of the transmission coefficient $i\mathrm{Im}({\ln}t)$ is a phase accumulated along the thickness $L$ of the system. In case of small phases, this approach can be correct if the system is immersed into a medium having averaged permittivity that compensates for the boundary effects. The immersion of the sample in the medium with averaged permittivity reduces the reflections from the boundaries of the sample. It is helpful in the case of long wavelength where the reduction of boundaries contribution allows consideration of smaller systems (in comparison to the vacuum case).

Let us note that the imaginary part of the complex logarithm (as well as the phase) is undefined with precision of $2\pi$. Later on, we consider the phase restored "by frequency". That means we choose the phase (for a fixed thickness of the system) in such a way that it increases monotonically alongside with the increasing frequency, and the phase of the transmitted signal at zero frequency is considered zero. This phase selection also corresponds to the phase recovered "by propagation" of the inhomogeneous sample, where it is supposed that the phase of the wave always increases in the direction of propagation. The identity of the methods is the result of consideration of the phase as a quantity that is continuous and increases monotonically with both the coordinate and frequency. For convenience, we refer to the phase restored "by frequency" as total phase, and the phase reduced to the region $[0,2\pi]$, as reduced phase.

The real part of ${\ln}|t|/L$ describes the decay of the field amplitude, which is determined by the imaginary part of the wave vector. Thus, the effective wave vector for a thick system can be defined as follows \cite{puzko2017analytical}:
\begin{equation}
\mathrm{Re}k_{eff}=\frac{\mathrm{Im}({\ln}t)}{L},\label{k1}
\end{equation}
\begin{equation}
\mathrm{Im}k_{eff}=-\frac{\mathrm{Re}({\ln}t)}{L}.\label{k2}
\end{equation}
These formulas can be combined into
\begin{equation}
	\label{keff}
	k_{eff}=-i\frac{{\ln}t}{L}.
\end{equation}

The definition of $k_{eff}$ by means of transmission coefficient is still valid for periodic systems that was confirmed by numerical calculations\cite{puzko2017analytical,vinogradov2002problem}.

\begin{figure}
	\centering\includegraphics[width=1\linewidth]{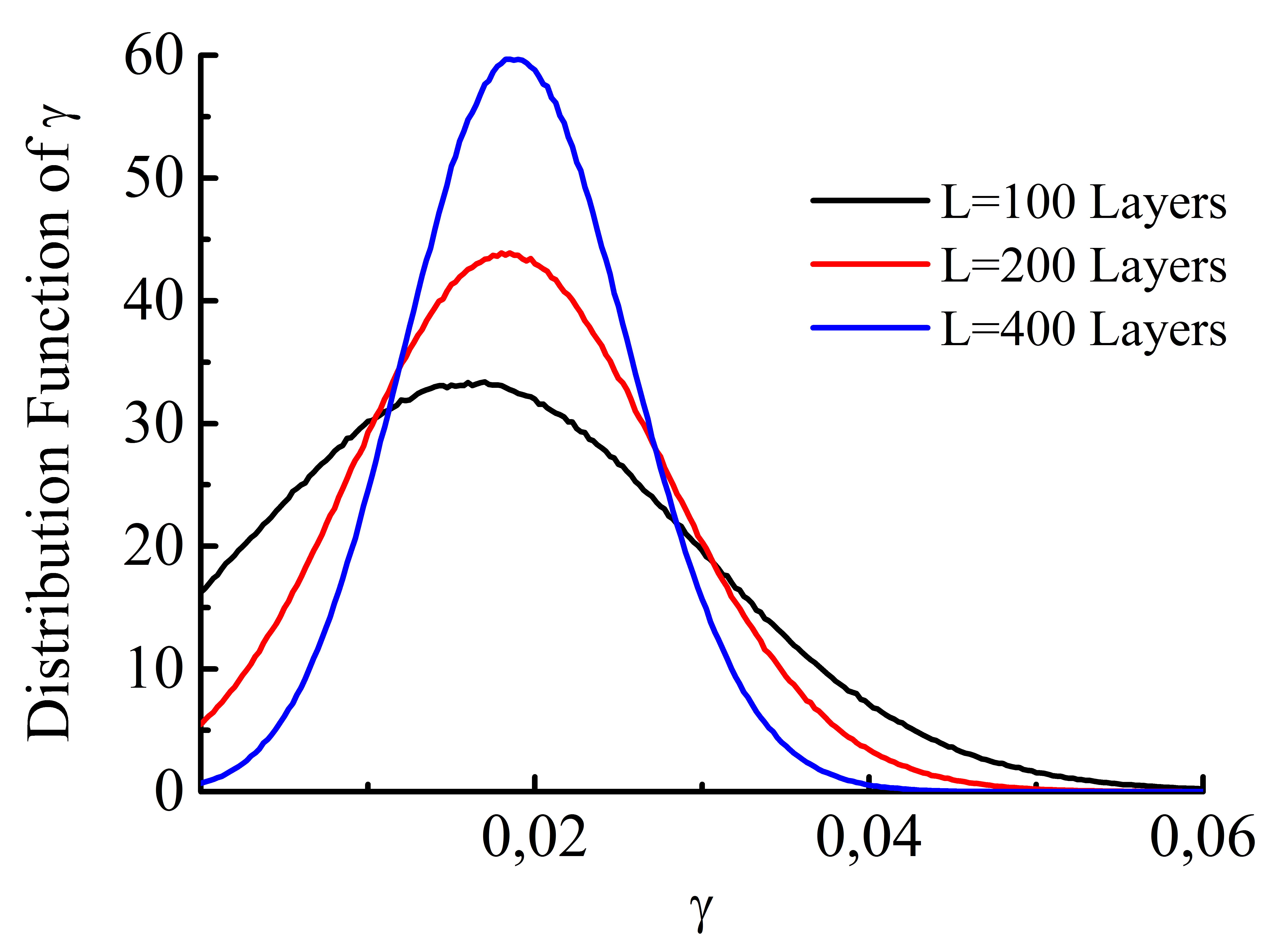}
	\caption{Distribution of $\gamma$ for random systems composed of 100 (black), 200 (red) and 400 (blue) layers. Dielectric permittivity of layers is uniformly distributed over the interval $[1;11]$. The thickness of each layer is $k_0d=1$. The ensembles used in the calculations consist of $10^7$ realizations.\label{gamma}}
\end{figure}

By definition (\ref{keff}), the imaginary part of the wave vector coincides with the inverse length of the localization, the Lyapunov index. That is, it self-averages, and its distribution is asymptotically Gaussian \cite{sheng2006introduction} (see Fig.~\ref{gamma}). The existence of the Anderson localization scale - the Anderson localization length - is equivalent to the self-averaging of $\mathrm{Im}k_{eff}$.

\begin{figure}[ht]
	\centering\includegraphics[width=1\linewidth]{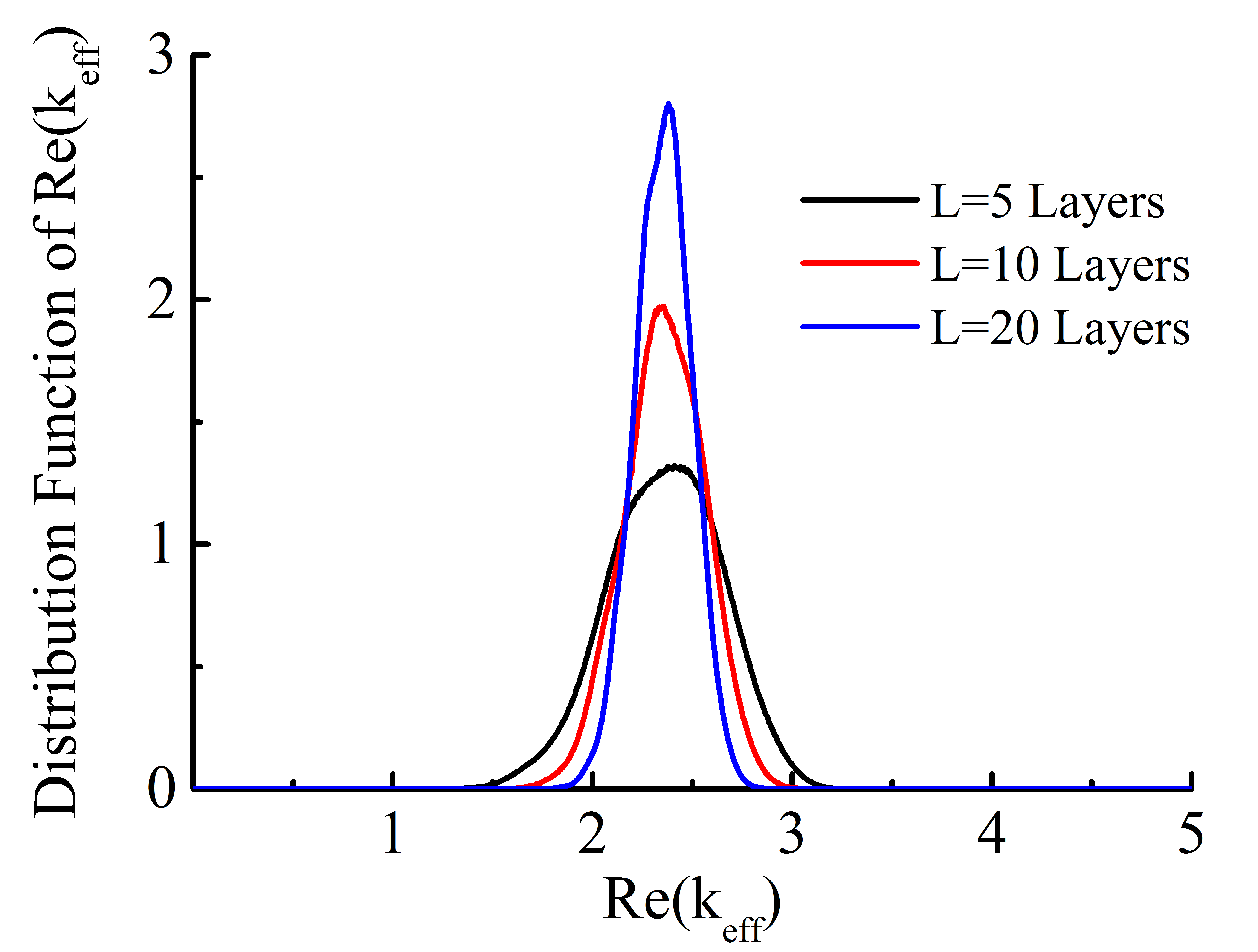}
	\caption{Distribution of $\mathrm{Re}k_{eff}$ for random systems composed of 100 (black), 200 (red) and 400 (blue) layers. The parameters are the same as in Fig.~\ref{gamma}.\label{Rekeff}}
\end{figure}

Let us now consider the real part of the effective wave vector. Calculations show that the real part demonstrates similar behavior (Fig.~\ref{Rekeff}).

\begin{figure}
	\centering\includegraphics[width=1\linewidth]{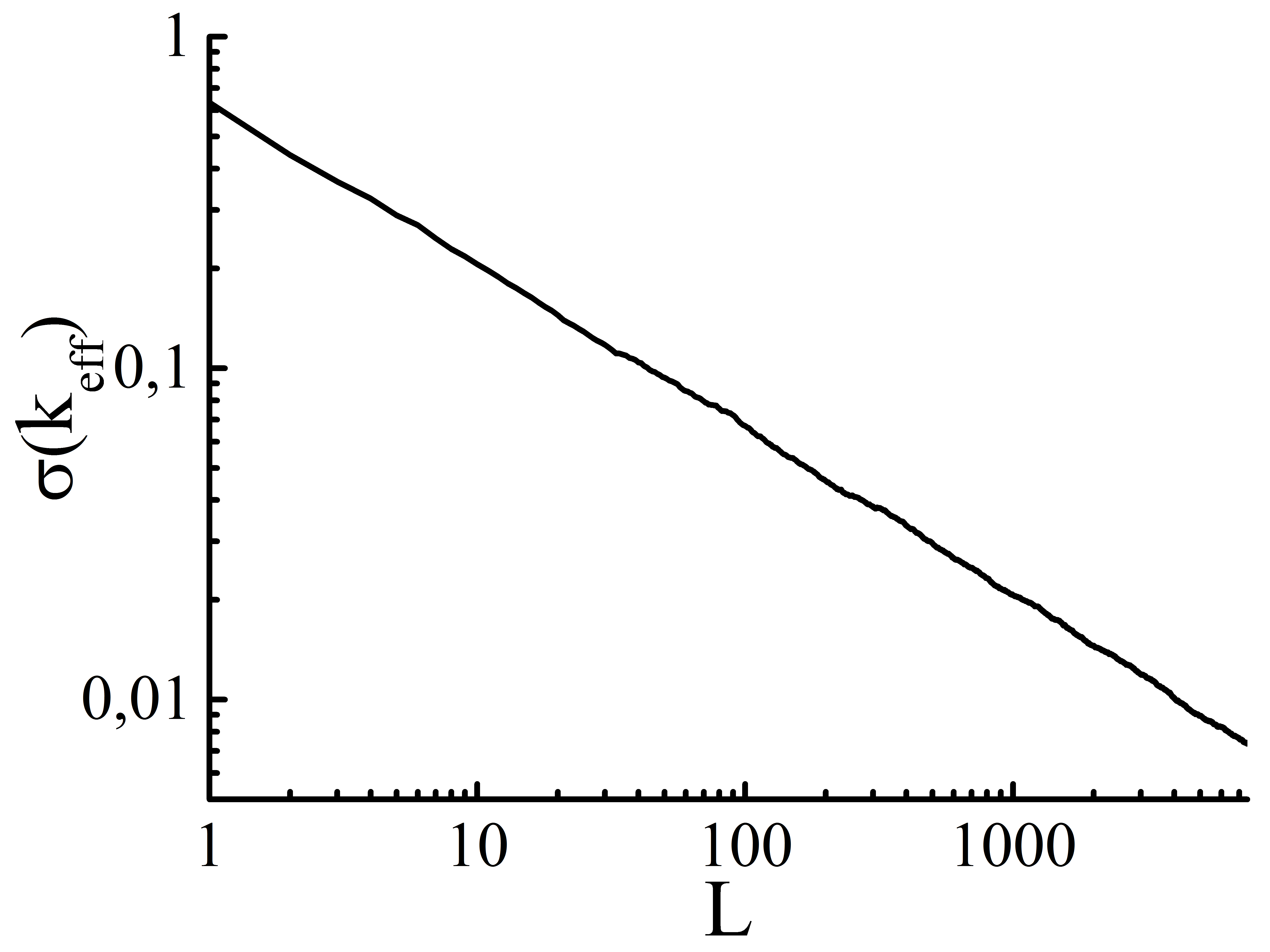}
	\caption{The dependency of variance of $\mathrm{Re}(k_{eff})$ distribution on the number of layers. Dielectric permittivity of layers is uniformly distributed over the interval $[1;11]$. The thickness of each layer is $k_0d=1$. The ensembles used in the calculations consist of $10^3$ realizations.\label{variance}}
\end{figure}

As can be seen from the graphs in Fig.~\ref{gamma}-\ref{Rekeff} for the real and imaginary parts of the effective wave vector, self-averaging of the effective wave vector occurs as the system thickness increases, i.e. the distribution function becomes narrower. The rate of self-averaging can be estimated. The variance of the distribution function as a thickness function is shown in Fig.~\ref{variance}. As one can see, the dependency of variance on the system length is a power law $\sigma\left(k_{eff}\right)\sim L^{-\alpha}$. The exponent obtained from the calculations is $\alpha\approx{0.5}$. Thus, one can expect that the effective wave vector will reach a certain value for a sufficiently long system. In other words, both the imaginary and the real parts of the effective wave vector self-average.

Thus, we have shown that $k_{eff}$ determined by the formula (\ref{keff}) self-averages as the thickness of the system increases not only in the long-wave approximation but for any ratio of the wavelength to the thickness of the layer. At the same time, the effective wave vector acquires an imaginary part due to wave localization in the system. It should be noted, that the number of the layers is undoubtedly important for self-averaging (the quantity averaging over the few layers is doubtful statement). However, the self-averaging occurs for any inhomogeneity size for sufficiently large (composed of many layers) system.

Now let us compare the definition (\ref{keff}) with other definitions of the effective wave vector. Numerical calculations for an ensemble of disordered systems shows that in the long-wave approximation the value of the effective wave vector calculated by the formula (\ref{k1}) coincides with that obtained from R-T retrieval (\ref{k_eff_Soukalis}). The frequency dependencies of averaged over the ensemble values of $k_{eff}$ in the long-wave limit for both definitions are close to the function $k_{st}=\sqrt{\langle\varepsilon\rangle}k_0$ (which corresponds to effective permittivity in static case), where $\varepsilon$ is the averaged over the distribution value of the permittivity. As it was mentioned earlier, the definition (\ref{keff}) is correct for small phases (in the approximation $\lambda\gg{L}$), when the system is immersed into a medium having averaged permittivity. Therefore, we assumed in the calculations that the admittance of the external medium is $Y_{av}=\sqrt{\langle\varepsilon\rangle}$, where the averaging is performed over the $\varepsilon$ distribution. Utilizing the external medium with this value of admittance makes it possible to reduce reflections from the boundaries of the system and, thus, reduces the dependency of the transmission coefficient on the boundaries. At the same time, the introduction of an admittance differing from the one of vacuum is not essential if the size of the system is significantly large in comparison with the wavelength.

The definition of the effective wave vector was introduced in \cite{vinogradov2004band} in terms of the Bloch wave vector $k_{Bl}$ of the specific photonic crystal, whose period is the considered layered sample. It was shown in \cite{vinogradov2004band} that the convergence $\left(\mathrm{Im}\langle{k_t}\rangle-\langle{\mathrm{Im}k_{Bl}}\rangle\right)_{Sp>2}\xrightarrow[L\to{\infty}]{}0$ takes place, where $k_t$ is the effective wave vector defined as (\ref{keff}), but since the measure of band gaps tends to unity (as also shown in \cite{vinogradov2004band}), then $\mathrm{Im}\langle{k_t}\rangle-\langle{\mathrm{Im}k_{Bl}}\rangle\xrightarrow[L\to{\infty}]{}0$. Thus, the convergence $\mathrm{Im}\langle{k_t}\rangle-\langle{\mathrm{Im}k_{Bl}}\rangle\xrightarrow[L\to{\infty}]{}0$ takes place for imaginary parts of the difference between differently defined effective vectors. Here we show that the convergence also takes place for real parts.
\begin{figure}
	\begin{minipage}{0.49\linewidth}
		\centering\includegraphics[width=1\linewidth]{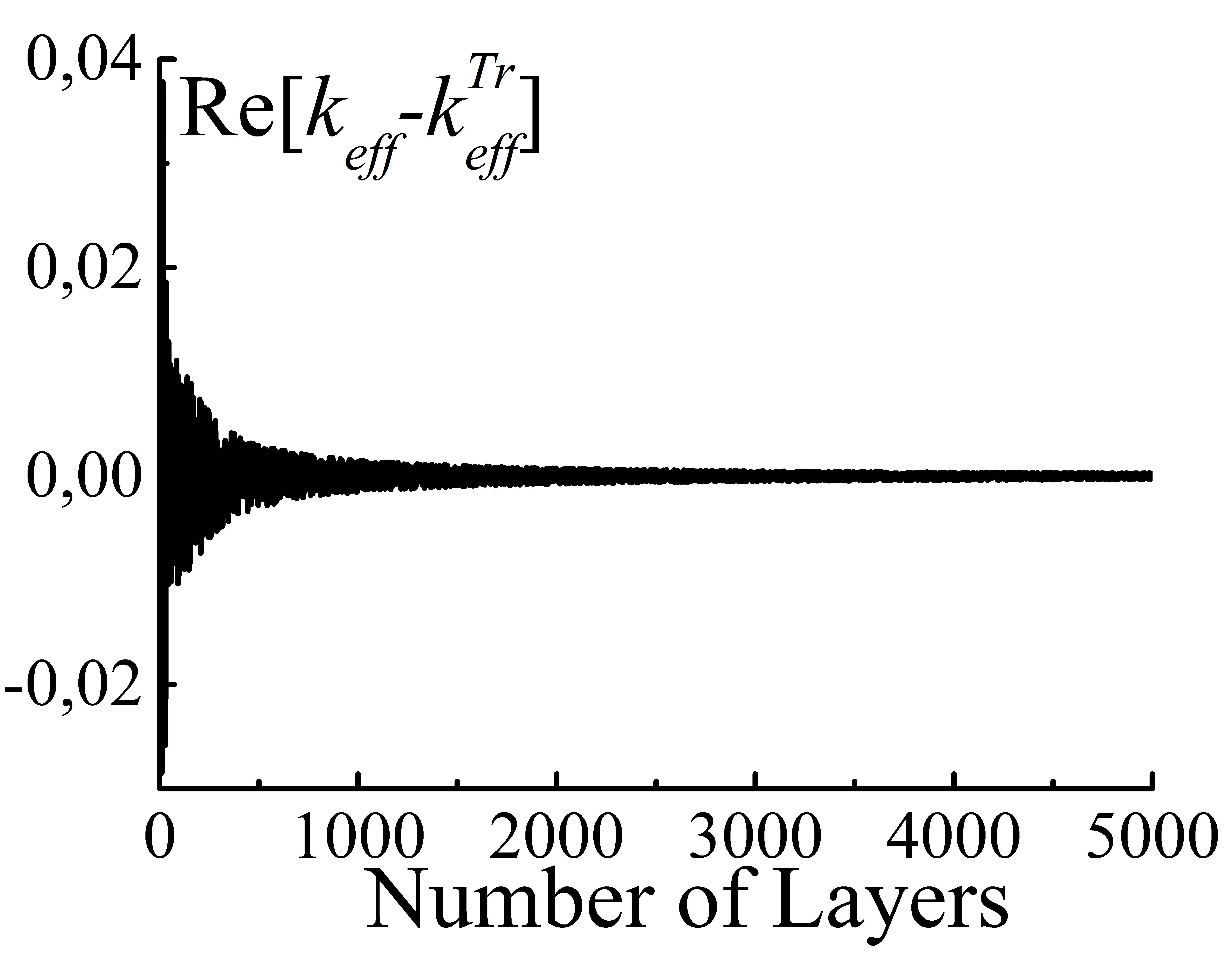}
		(a)
	\end{minipage}
	\begin{minipage}{0.49\linewidth}
		\centering\includegraphics[width=1\linewidth]{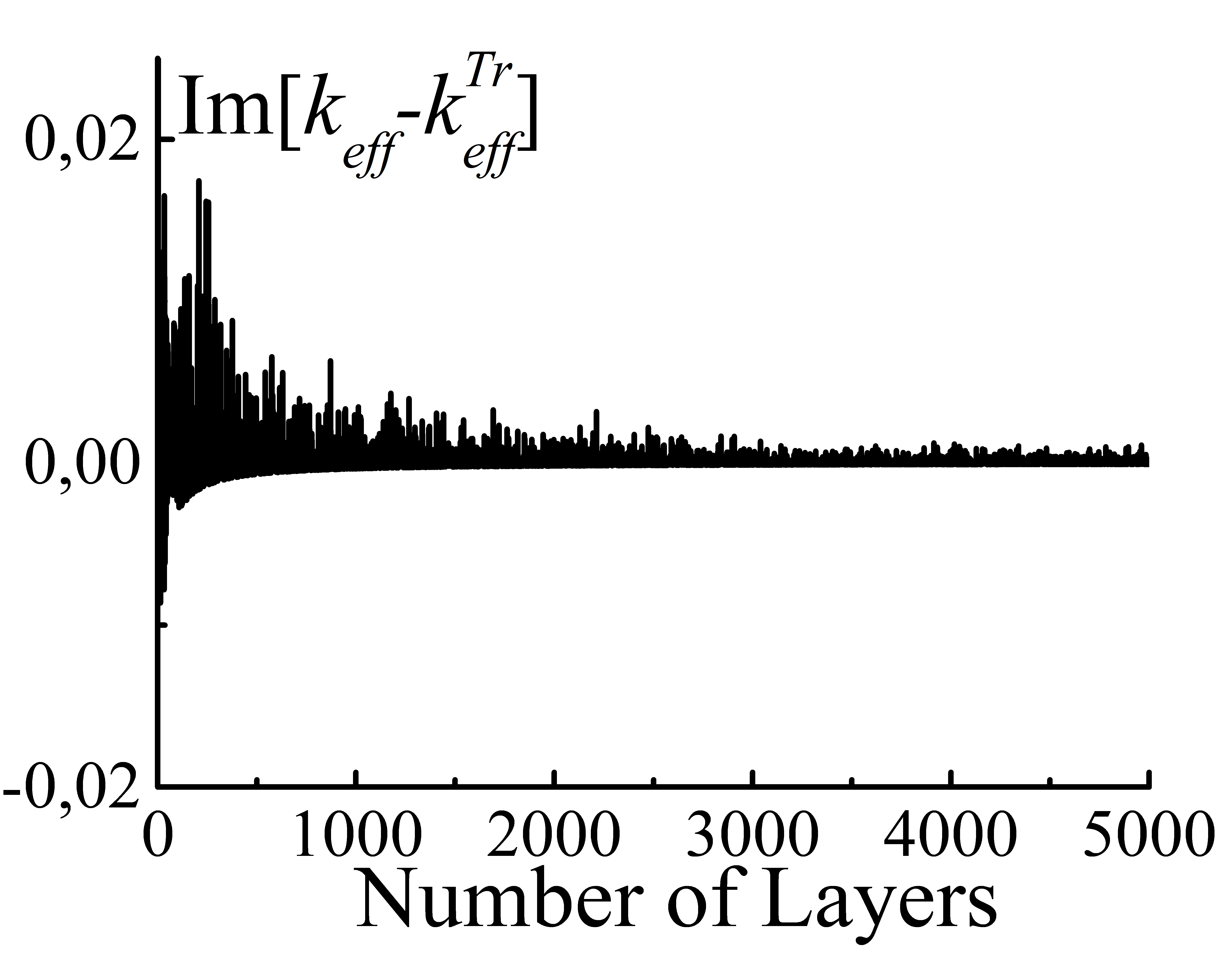}
		(b)
	\end{minipage}
	\caption{The dependencies for (a) real and (b) imaginary parts of $k_{eff}-k_{eff}^{Tr}$ on the thickness of a random system. Dielectric permittivity of layers is uniformly distributed over the interval $[1;11]$. The graphs represent the dependency for a single system realization for each thickness.\label{kBl}}
\end{figure}

The results of numerical calculations of $\mathrm{Re}(k_t-k_{Bl})$ are depicted in Fig.~\ref{kBl}. The calculations are made for a single realization of the system of each thickness. It can be seen that the difference between the two definitions of the effective wave vector decreases as the system size increases. It should be noted that the real part of $k_{Bl}$ frequently changes with the thickness of the system by jumps of value $\sim\pi/L$. It happens due to the fact that the measure of band gaps tends to unity as the thickness increases, i.e. the Bloch wave vector must correspond to the band gap at the limit $L\to\infty$. Despite the fact that the phase shift on the period of the considered photonic crystal changes abruptly, $k_{Bl}$ and $k_t$ tend to the same value.

Thus, the effective wave vector introduced by (\ref{keff}) corresponds to the definitions given in \cite{smith2002determination} and \cite{vinogradov2004band}.

\section{Qualitative arguments on the self-averaging of the effective wave vector}
Following by Sheng \cite{sheng2006introduction}, let us consider two large (consisting of many layers and with a total thickness significantly larger than the wavelength) pieces of a 1D disordered system referred to as pieces A and B. The transmission coefficients of A and B are $t_A$ and $t_B$, respectively. The system composed of these two pieces has the transmission coefficient $t_{AB}$ which equals (in order of magnitude) to $t_{AB}\approx{t_A}{t_B}$. Now, adding one more piece referred to as C, we assume that $t_{ABC}\approx{t_{AB}t_C}\approx{t_A}{t_B}{t_C}$, i.e. with logarithmic accuracy
\begin{equation}
\ln{t_{ABC}}\approx\ln{t_A}+\ln{t_B}+\ln{t_C}
\label{3pieces}
\end{equation}
This relation is valid not only for modules, but also for phases. However, that is true only if we choose the phases reconstructed "by propagation", that is, the phase of the wave is supposed to be an increasing function of system thickness. As mentioned previously, this method of full phase reconstruction is equal to the reconstruction "by frequency".

Since $\ln{t_i}$ are independent and identically distributed, then $\frac{1}{N}\sum_{i=1}^{N}\ln{t_i}$ is asymptotically Gaussian distributed in accordance with the law of large numbers. The variance of the distribution is $\sigma\sim{1/\sqrt{N}}$, i.e. $D=\sigma^2\sim{1/L}$ as soon as all the layers of the system are of the same thickness. That is why the quantities $\phi/L$ and $\gamma$ are self-averaging and, moreover, $D_\gamma\sim{D_k}\sim 1/L$, which completely corresponds to the numerical experiment (see Fig.~\ref{variance}).

\section{Frequency dependence of real and imaginary parts on the effective wave vector}
\begin{figure}
	\begin{minipage}{0.49\linewidth}
		\centering\includegraphics[width=1\linewidth]{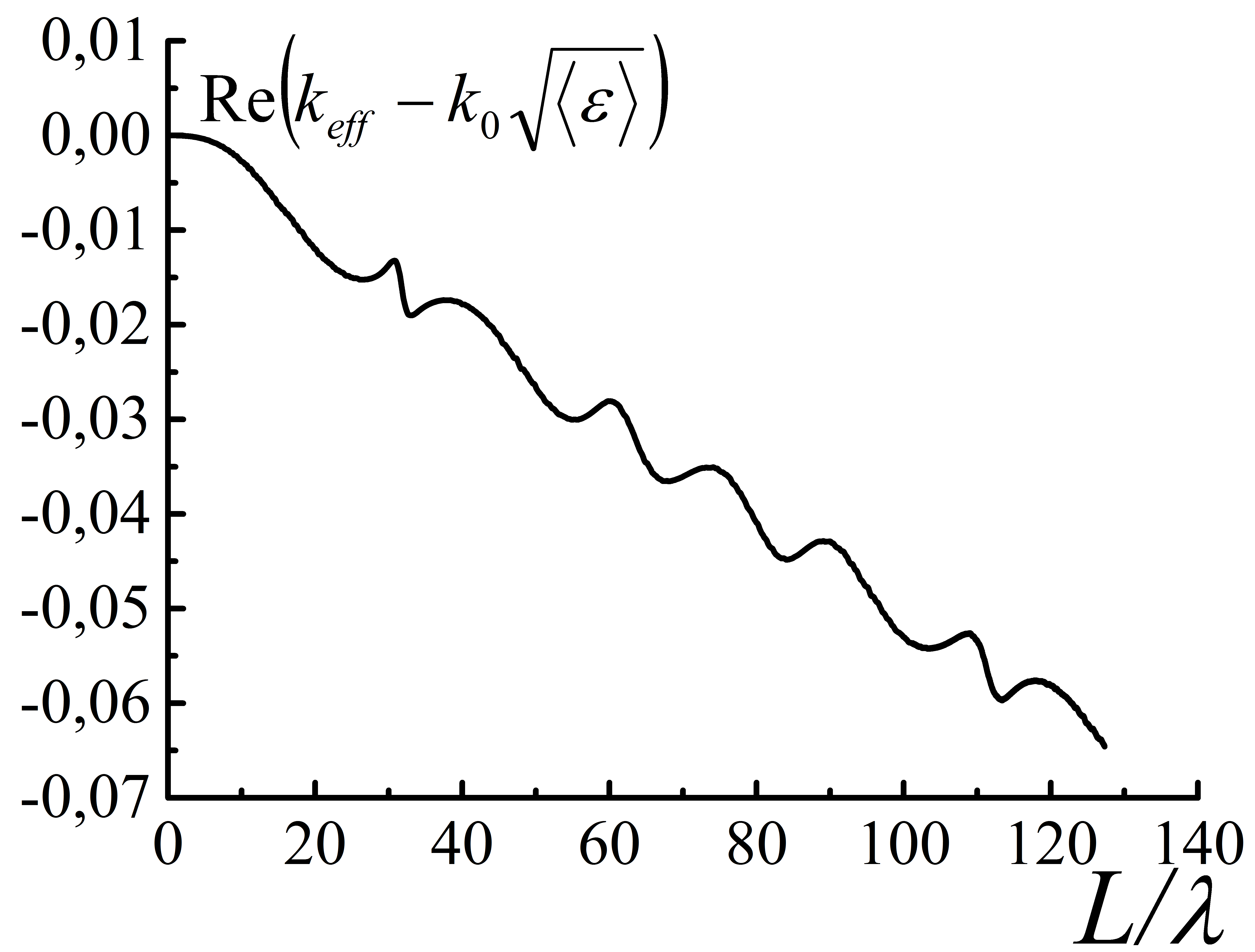}
		(a)
	\end{minipage}
	\begin{minipage}{0.49\linewidth}
		\centering\includegraphics[width=1\linewidth]{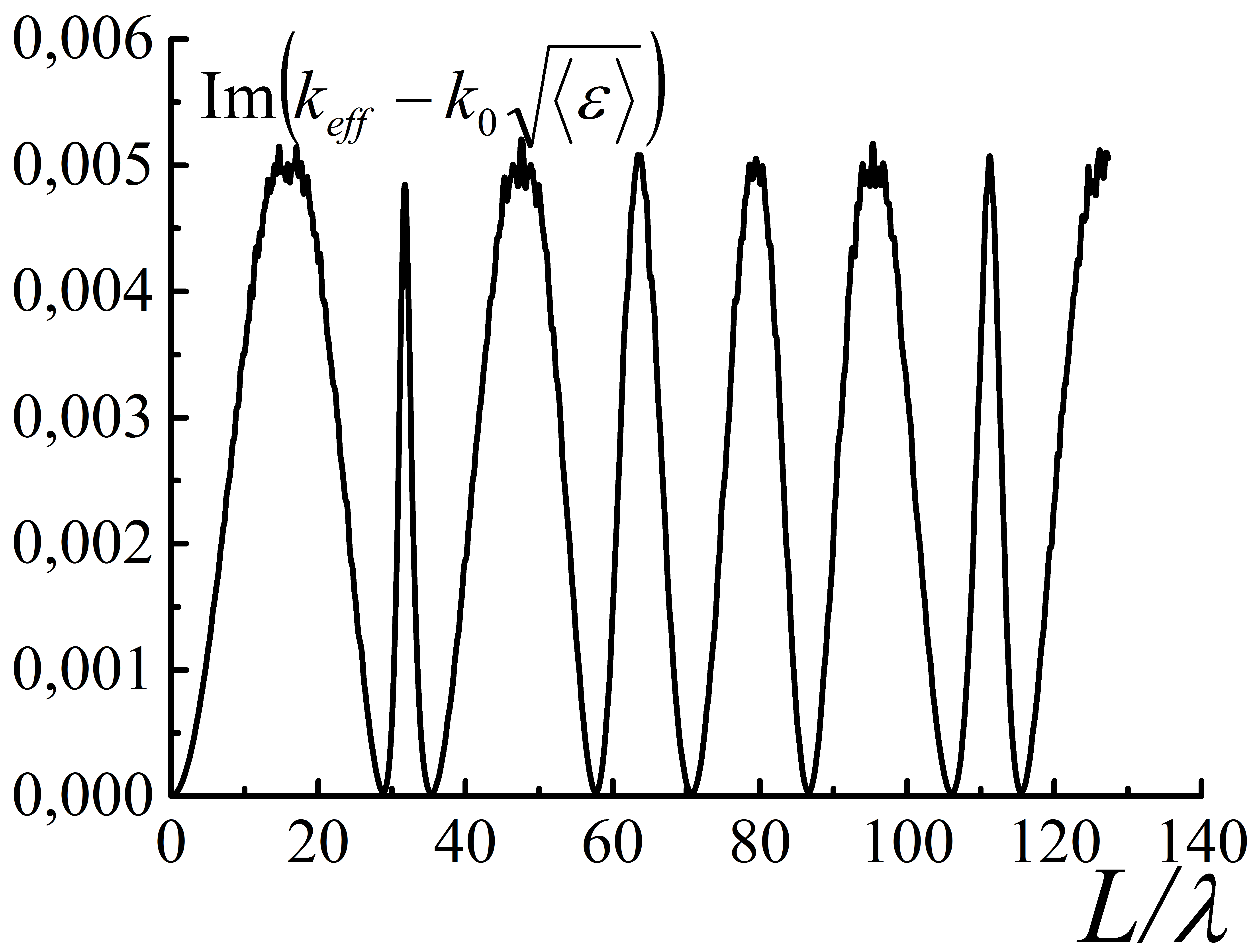}
		(b)
	\end{minipage}
	\caption{Frequency dependencies of (a) real and (b) imaginary parts of $k_{eff}-k_0\sqrt{\langle\varepsilon\rangle}$ averaged over ensemble of random systems. Each system is composed of 100 layers. The ensemble consists of 40000 realizations. Dielectric permittivity of each layer is taken to be $\varepsilon_1=2$ or $\varepsilon_2=3$ with equal probabilities.\label{k_w}}
\end{figure}
	
\begin{figure}
	\begin{minipage}{0.49\linewidth}
		\centering\includegraphics[width=1\linewidth]{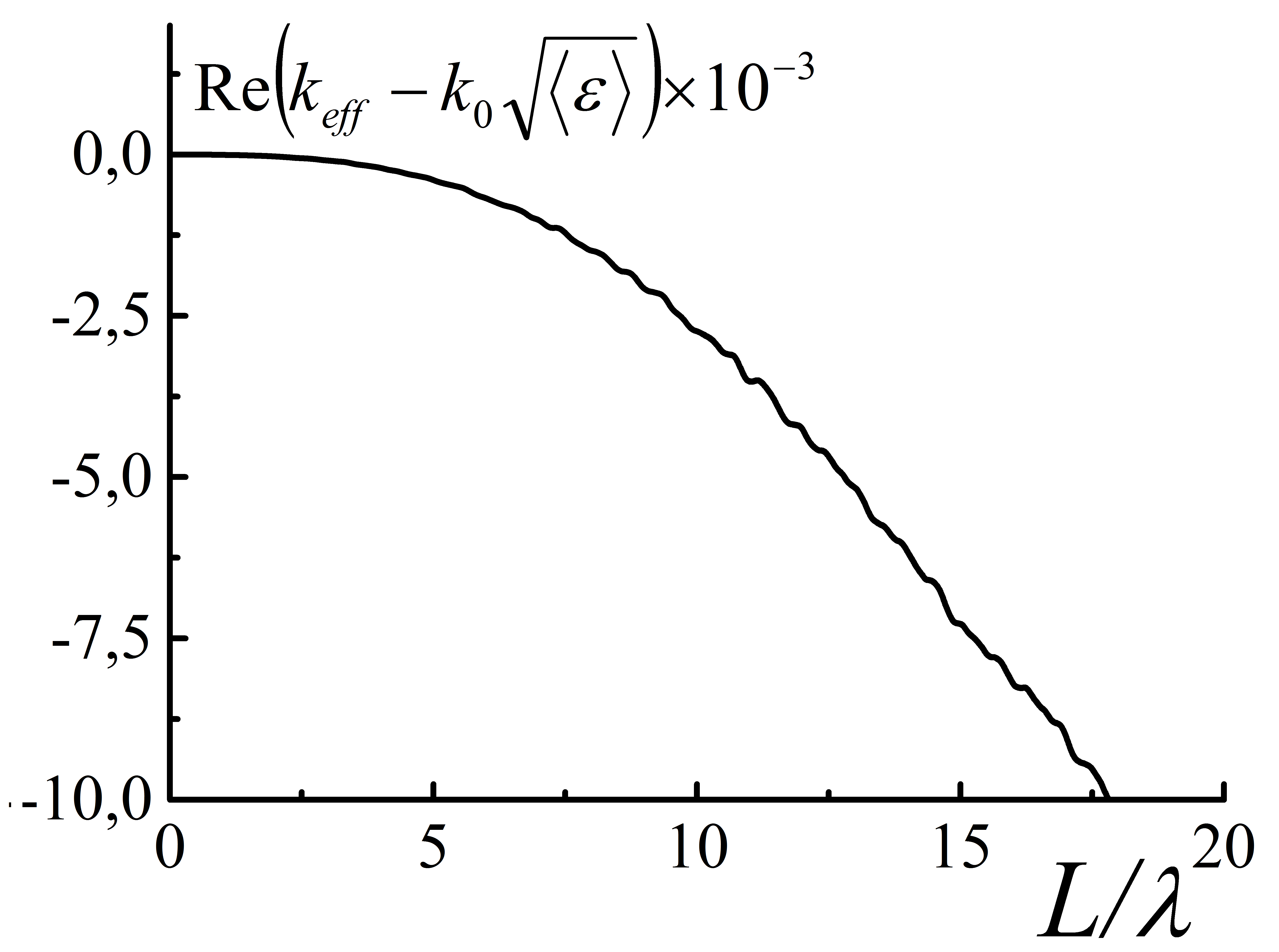}
		(a)
	\end{minipage}
	\begin{minipage}{0.49\linewidth}
		\centering\includegraphics[width=1\linewidth]{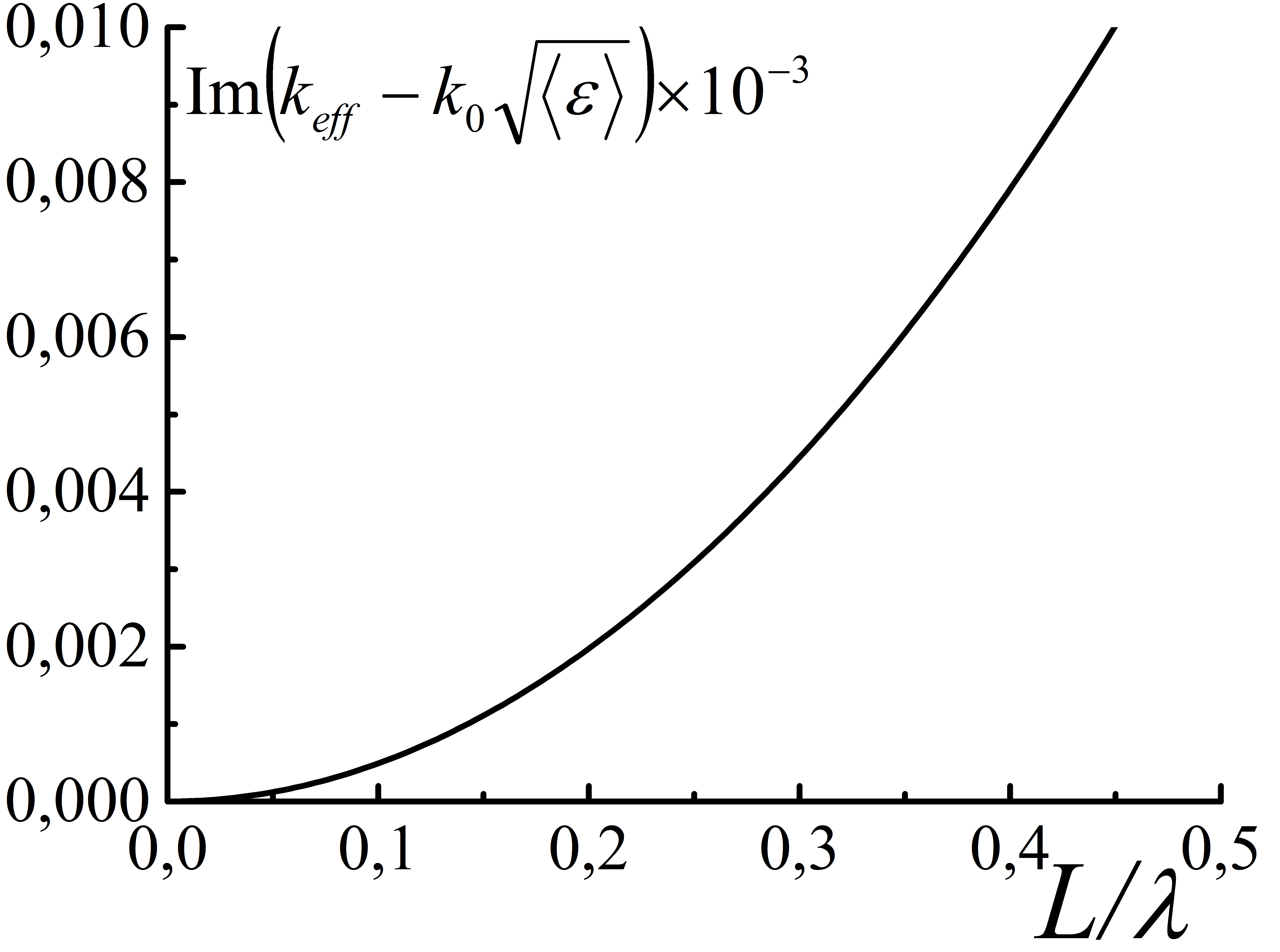}
		(b)
	\end{minipage}
	\caption{Frequency dependencies of (a) real and (b) imaginary parts of $k_{eff}-k_0\sqrt{\langle\varepsilon\rangle}$ averaged over ensemble of random systems. Each system is composed of 100 layers. The ensemble consists of 40000 realizations. Dielectric permittivity of each layer is taken to be $\varepsilon_1=2$ or $\varepsilon_2=3$ with equal probabilities.\label{k_w_large}}
\end{figure}

Consider the dispersion properties of the introduced effective wave vector. The graphs of the dependencies are shown in Fig.~\ref{k_w}-\ref{k_w_large} for layered systems composed of two types of layers. It can be seen that the effective wave vector is close to the value $k_{st}=\sqrt{\langle\varepsilon\rangle}k_0$ in the long-wavelength limit. However, as the frequency increases, the dependency changes and becomes $k_{eff}\approx\langle\varepsilon\rangle{k_0}$. The imaginary part of $k_{eff}$ turns to 0 in some points, which is due to the fact that one of the layers becomes transparent at these frequencies, i.e. its transmission coefficient equals to $\pm 1$. The dispersion dependency (see Fig.~(\ref{k_w_large})) of the imaginary part of $k_{eff}$ behaves as $\mathrm{Im}k_{eff}\sim\lambda^{-2}$ in the long-wavelength approximation, which is in accordance with the known behavior of the Lyapunov index.

\section{Kramers-Kronig like relations for effective wave vector}
Since both $k=\mathrm{Re}k_{eff}$ and $\gamma=\mathrm{Im}k_{eff}$ are self-averaging quantities and have a clear physical meaning, a possible connection between them is of a peculiar interest.

We examine this question by means of the transfer matrix method \cite{yariv1984optical} (transfer matrix denoted by capital letter in the paper in contrast to transmission coefficient denoted by small letter). It is easy to see that the transmission coefficient is an analytic function of the frequency: the T-matrix of the whole system can be written in terms of the transmission and reflection coefficients
\begin{equation}
T=\begin{pmatrix}
t-\frac{r_Lr_R}{t} & -\frac{r_L}{t}\\
\frac{r_R}{t}	&	\frac{1}{t}
\end{pmatrix}
\label{Tmatrix}
\end{equation}
where $r_L$ and $r_R$ are reflection coefficients from both the left and right sides of the system, respectively (the transmission coefficient is independent from the direction). On the other hand, the T-matrix of the whole system is the product of T-matrices of homogeneous layers forming the system. All elements of these T-matrices are analytic functions of frequency. Thus, the transmission coefficient, being a fractionally-rational function of these elements, is also an analytic function of the frequency. Consequently, the function
\begin{equation}
\frac{1}{t}\frac{dt}{d\omega}=\frac{d\ln(t)}{d\omega}=\frac{dk_{eff}}{d\omega}
\label{analytic}
\end{equation}
should also be analytical. Therefore, the following relations are valid
\begin{eqnarray}
	\label{KrKr1}
	\frac{d\mathrm{Re}k_{eff}(\omega)}{d\omega}=const+\frac{1}{\pi}v.p.\int_{-\infty}^{+\infty}\frac{\frac{d\mathrm{Im}k_{eff}(u)}{du}}{u-\omega}du,\\
	\label{KrKr2}\frac{d\mathrm{Im}k_{eff}(\omega)}{d\omega}=-\frac{1}{\pi}v.p.\int_{-\infty}^{+\infty}\frac{\frac{d\mathrm{Re}k_{eff}(u)}{du}}{u-\omega}du.
\end{eqnarray}
The Kramers-Kronig like relations (\ref{KrKr1}) and (\ref{KrKr2}) are well-known property of optical parameters connected with the causality principle~\cite{bookKramKron}. With respect to the definition of the Lyapunov exponent $\gamma(\omega)$, the second relation can be rewritten as (we made the notation $k=\mathrm{Re}k_{eff}$)
\begin{equation}
\frac{d\gamma(\omega)}{d\omega}=-\frac{1}{\pi}v.p.\int_{-\infty}^{+\infty}\frac{\frac{dk(u)}{du}}{u-\omega}du
\label{eq12}
\end{equation}

Taking into account parity $k(-\omega)=-k(\omega)$ (hence, $dk/d\omega$ is an even function of frequency) and utilizing relation between the wave vector and the density of states ($\rho(\omega)=dk/d\omega$), one can get
\begin{equation}
\frac{d\gamma(\omega)}{d\omega}=-\frac{2}{\pi}v.p.\int_{0}^{+\infty}\frac{\omega\rho(u)}{u^2-\omega^2}du
\label{eq13}
\end{equation}

Performing the integration over the frequency, one obtains the Johnes-Huberd-Thouless formula \cite{thouless1972relation,herbert1971localized}
\begin{equation}
\gamma(\omega)=\frac{1}{\pi}v.p.\int_{0}^{+\infty}\rho(u)\ln\left|1-\frac{\omega^2}{u^2}\right|du
\label{JHT}
\end{equation}

Thus, the frequency derivative of the real part of the effective wave vector corresponds to the density of states, which confirms once again that $k_{eff}$ has a physical ground to be considered as an effective wave vector.

\section{Phase randomization}
It was shown in the preceding sections that the effective wave vector self-averages. Let us now consider in detail the behavior of phase~(\ref{t_represent}), which is directly related to the effective wave vector for each system realization as $\phi=-i\ln\left(t/|t|\right)=\mathrm{Im}\ln{t}=k_{eff}L$. Despite the fact that $k_{eff}$ self-averages and, as a consequence, is a deterministic quantity, the phase randomizes; more precisely, the reduced phase, which is a phase reduced to an interval $[0,2\pi]$, randomizes. Phase randomization is used as basis for creating a random T-matrix approach \cite{anderson1980new,lambert1983random,stone1983phase,izrailev1998classical}.

The apparent contradiction can be easily explained as one notes that variance of the wave vector tends to zero as $D_k=\sigma_k^2\sim 1/L$ and the variance of the phase increases as $D_\phi=L^2\sigma_k^2\sim{L}$. Since the distribution of $k_{eff}$ is similar to the Gaussian distribution (see Fig.~\ref{Rekeff}), the distribution for the total phase is expected to be similar to Gaussian one, but as the system thickness increases, the distribution shifts and gets wider.
\begin{figure}
	\centering\includegraphics[width=1\linewidth]{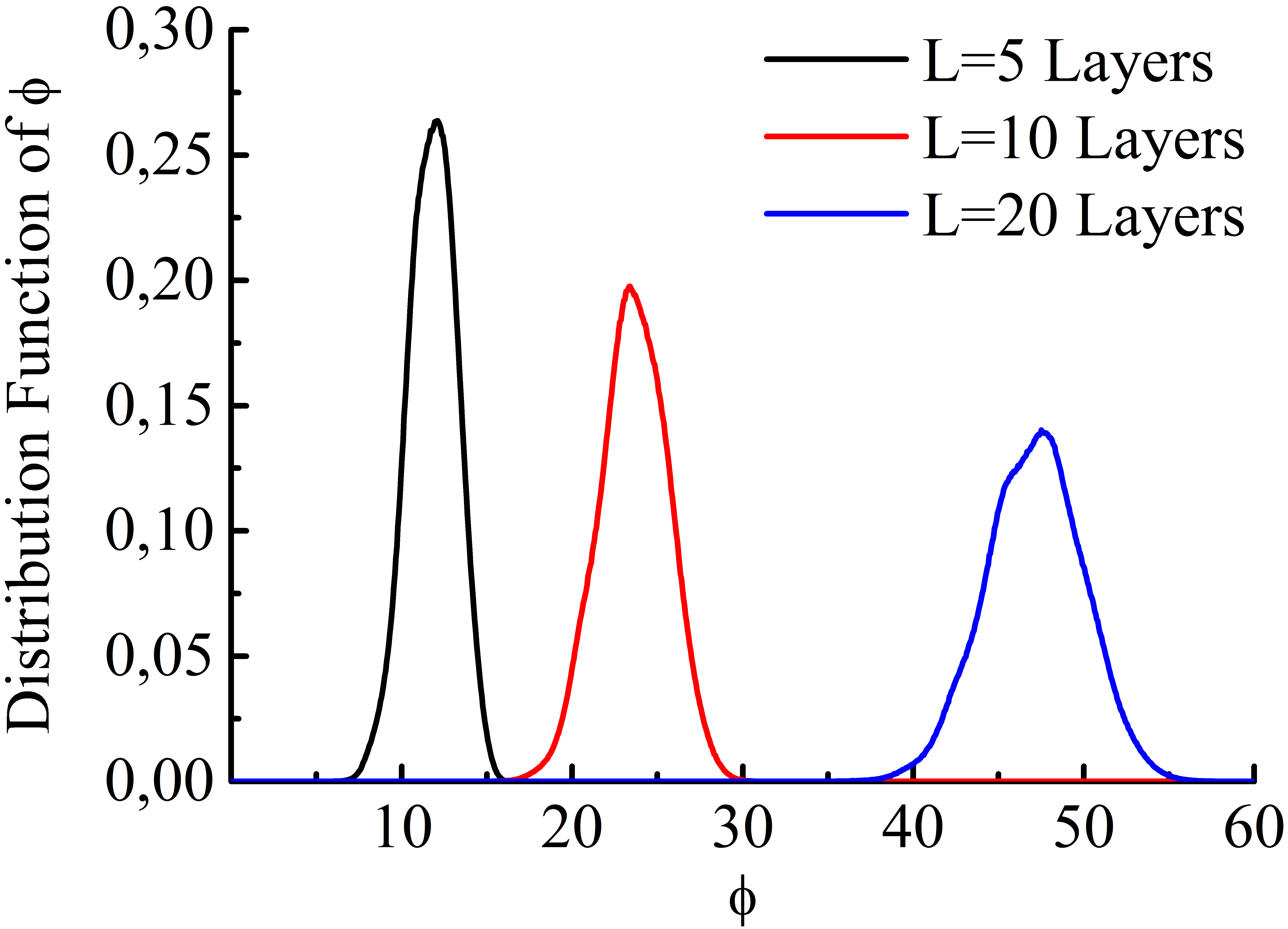}
	\caption{Distribution of $\phi$ for random systems composed of 5 (black), 10 (red) and 20 (blue) layers. The parameters are the same as in Fig.~\ref{gamma}.\label{phase}}
\end{figure}

\begin{figure}[ht]
	\centering\includegraphics[width=1\linewidth]{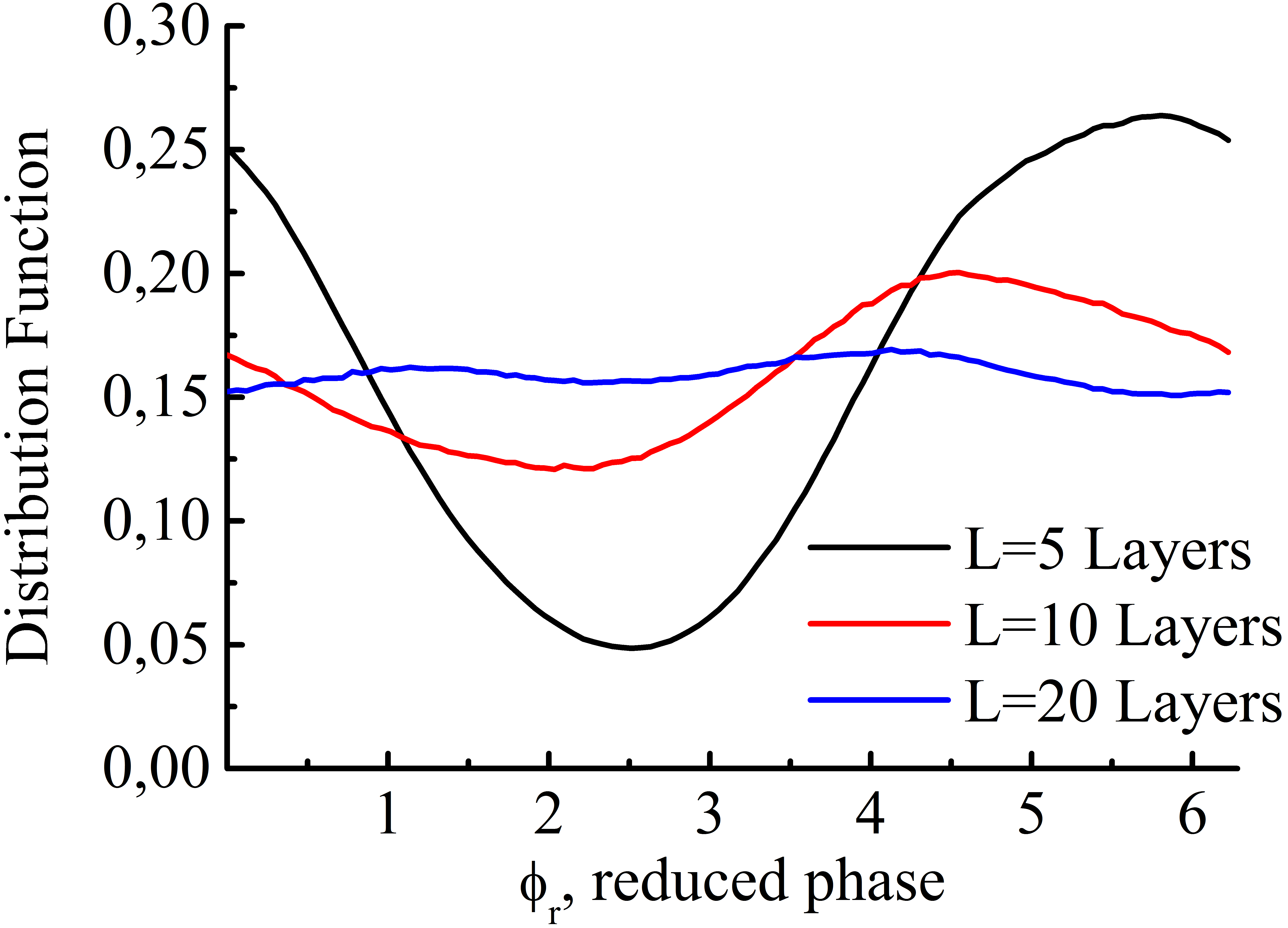}
	\caption{Distribution of reduced phase for random systems composed of 5 (black), 10 (red) and 20 (blue) layers. The parameters are the same as in Fig.~\ref{gamma}.\label{reduced_phase}}
\end{figure}

The numerical calculations confirm this behavior (Fig.~\ref{phase})\footnote{It is important that our system is placed not in vacuum but in a medium with permittivity averaged over $\varepsilon$ distribution. The scattering at the "vacuum-medium" boundary introduces an additional shift to the phase of the transmitted wave and changes the shape of the graphs.}. The distribution shifts to the right proportionally to the system size and broadens. Therefore, after the reduction to the interval, the distribution of the reduced phase should be close to the constant for a sufficiently long system (Fig.~\ref{reduced_phase}).

However, even for a large thickness of the system, the calculations show that the distribution function differs from a constant. That is due to the presence of an additional correlation scale - the width of the layer.

\section{Conclusion}
In this paper, we considered the propagation of the electromagnetic wave through the disordered layered system. It is well-known that even in the longwave approximation the propagation of the wave cannot be described in terms of effective dielectric permittivity and/or magnetic permeability (because of mesoscopic nature of effective permittivity and permeability) \cite{vinogradov2011additional,vinogradov2001electrodynamic}. However, we have shown that the effective wavenumber self-averages and, thus, it is macroscopic quantity.

The imaginary part of the proposed effective wave vector is associated with the Lyapunov exponent (inverse localization length). To clarify the physical meaning of the real part of the introduced effective wave vector, it is reasonable to consider the frequency derivative of the effective wave vector. This quantity is an analytical function of frequency (because it can be directly expressed in terms of the propagation coefficient) and thus satisfies the Kramers-Kronig like relation. These relations lead to the Jones-Herbert-Thouless relation, connecting the spectral density and the localization of eigenstates, i.e. the real part of the introduced effective wave vector gives us the correct value of the density of states.

We have shown that the proposed effective wave vector is a self-averaging quantity, i.e. it tends to a constant value as the system length (number of layers) increases. The self-averaging of $\mathrm{Re}k_{eff}$ and $\mathrm{Im}k_{eff}$ occurs not only within the limit of long-wavelength approximation but for an arbitrary ratio between the wavelength and inhomogeneity size.

Moreover, the self-averaging of $k_{eff}$ describes the Anderson localization phenomenon, i.e. $\mathrm{Im}k_{eff}$ self-averages to the quantity that is Lyapunov exponent in the limit of $L\to\infty$. Thus, the homogenization of the effective wave vector allows the unified description of electromagnetic waves propagation and localization.

\section*{Acknowledgments}
We would like to thank prof. A.P. Vinogradov and prof. C.R. Simovski for helpful discussions. The work was supported by the Foundation for the Advancement of Theoretical Physics and Mathematics "BASIS" and by Russian Foundation for Basic Research (RFBR) (grant 18-52-00044).
%%%%%%%%%%%%%%%%%%%%%%% References %%%%%%%%%%%%%%%%%%%%%%%%%

%Add references with BibTeX or manually.
%\cite{Zhang:14,OSA,FORSTER2007,Dean2006}

%%%%%%%%%% If using BibTeX:
\bibliography{self}

\end{document}